\documentclass[a4paper, authorcolumns]{lipics-v2021}
\usepackage{mathtools}
\usepackage{cite}
\nolinenumbers

\newcommand{\RE}{\mathbb{R}}

\newcommand{\eps}{\varepsilon}

\newcommand{\bdOmega}{\partial \kern+1pt \Omega} 

\newcommand{\SP}{\kern+1pt}             
\newcommand{\revFunk}[1]{{}^r \kern-1pt F_{#1}}

\hideLIPIcs
\nolinenumbers

\title{Proximity Alert: Ipelets for Neighborhood Graphs and Clustering}
\titlerunning{Proximity Alert, Ipelets for Neighborhood Graphs and Clustering}

\author{Gitan Balogh}{Department of Computer Science, University of Maryland, College Park, USA \and \url{~}}{gitan@terpmail.umd.edu}{}{}

\author{June Cagan}{Department of Computer Science, University of Maryland, College Park, USA  \and \url{~}}{jcagan@terpmail.umd.edu}{}{}

\author{Bea Fatima}{Department of Physics, Utrecht University, Utrecht, The Netherlands \and \url{~}}{b.fatima@students.uu.nl}{}{}

\author{Auguste H. Gezalyan}{Université de Lorraine, CNRS, Inria, LORIA, F-54000 Nancy, France}{octavo@umd.edu}{https://orcid.org/0000-0002-5704-312X}{}

\author{Danesh Sivakumar}{Department of Computer Science, University of Maryland, College Park, USA \and \url{~}}{dsivakumar@terpmail.umd.edu}{https://orcid.org/0009-0008-2484-5549}{}

\author{Arushi Srinivasan}{Department of Computer Science, University of Maryland, College Park, USA \and \url{~}}{asrin708@terpmail.umd.edu}{}{}

\author{Yixuan Sun}{Department of Computer Science, University of Maryland, College Park, USA \and \url{~}}{ysun1221@umd.edu}{}{}

\author{Vahe Zaprosyan}{Department of Computer Science, University of Maryland, College Park, USA \and \url{~}}{vahezap@umd.edu}{}{}

\author{David M. Mount}{Department of Computer Science, University of Maryland, College Park, USA \and \url{https://www.cs.umd.edu/~mount/}}{mount@umd.edu}{https://orcid.org/0000-0002-3290-8932}{}

\authorrunning{Balogh, Cagan, Fatima, Gezalyan, Sivakumar, Srinivasan, Sun, Zaprosyan, Mount}

\Copyright{Gitan Balogh, June Cagan, Bea Fatima, Auguste H. Gezalyan, Danesh Sivakumar, Arushi Srinivasan, Yixuan Sun, Vahe Zaprosyan, and David M. Mount}


\category{Media Exposition}

\date{\today}

\supplementdetails[subcategory={Software}]{Software}{https://github.com/Otcavo/Ipelets-Full-Library}
\supplementdetails[subcategory={Video}]{Media}{https://youtu.be/NQr-SY1x2qc}

\EventEditors{Hee-Kap Ahn, Michael Hoffmann, and Amir Nayyeri}
\EventNoEds{3}
\EventLongTitle{42nd International Symposium on Computational Geometry (SoCG 2026)}
\EventShortTitle{SoCG 2026}
\EventAcronym{SoCG}
\EventYear{2026}
\EventDate{June 2--5, 2026}
\EventLocation{New Brunswick, NJ, USA}
\EventLogo{socg-logo}
\def\EventLogoHeight{42}
\SeriesVolume{367}
\ArticleNo{108}  

\begin{document}

\maketitle

\begin{abstract}
Neighborhood graphs and clustering algorithms are fundamental structures in both computational geometry and data analysis. Visualizing them can help build insight into their behavior and properties. The Ipe extensible drawing editor, developed by Otfried Cheong, is a widely used software system for generating figures. One particular aspect of Ipe is the ability to add Ipelets, which extend its functionality. Here we showcase a set of Ipelets designed to help visualize neighborhood graphs and clustering algorithms. These include: $\eps$-neighbor graphs, furthest-neighbor graphs, Gabriel graphs, $k$-nearest neighbor graphs, $k^{th}$-nearest neighbor graphs, $k$-mutual neighbor graphs, $k^{th}$-mutual neighbor graphs, asymmetric $k$-nearest neighbor graphs, asymmetric $k^{th}$-nearest neighbor graphs,  relative-neighbor graphs, sphere-of-influence graphs, Urquhart graphs, Yao graphs, and clustering algorithms including complete-linkage, DBSCAN, HDBSCAN, $k$-means, $k$-means++, $k$-medoids, mean shift, and single-linkage. Our Ipelets are all programmed in Lua and are freely available.
\end{abstract}

\section{Introduction}
\begin{figure}[htbp]
    \centerline{\includegraphics[scale=0.85]{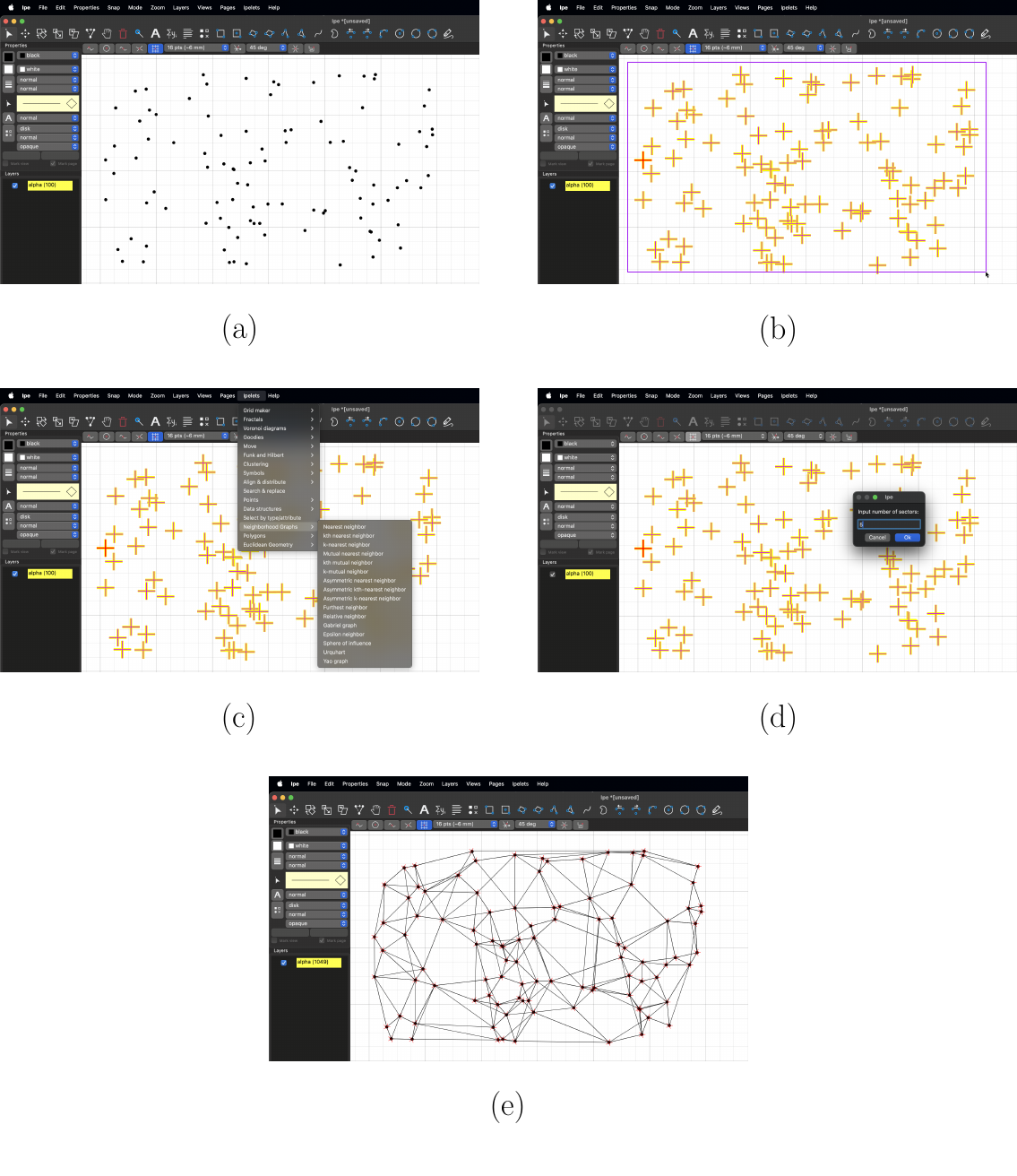}}
    \caption{(a) Place a point set in Ipe, (b) select the points, (c) choose a graph from the  ``Ipelets'' menu, (d) enter a parameter value if prompted, (e) the resulting graph is drawn automatically.}
    \label{fig:instructions}
\end{figure}

We present a set of \textbf{Ipelets} for visualizing clustering and neighborhood graph algorithms on point sets in $\RE^2$. These Ipelets include: $\eps$-neighbor graphs, furthest-neighbor graphs, Gabriel graphs, $k$-nearest neighbor graphs, $k^{th}$-nearest neighbor graphs, $k$-mutual neighbor graphs, $k^{th}$-mutual neighbor graphs, asymmetric $k$-nearest neighbor graphs, asymmetric $k^{th}$-nearest neighbor graphs,  relative-neighbor graphs, sphere-of-influence graphs, Urquhart graphs, Yao graphs, and clustering algorithms including complete-linkage, DBSCAN, HDBSCAN, $k$-means, $k$-means++, $k$-medoids, mean shift, and single-linkage. Some of these are also implemented in CGAL \cite{fabri2009cgal}. All our Ipelets are programmed in Lua and are freely available on \href{https://github.com/Otcavo/Ipelets-Full-Library}{GitHub}. To install any Ipelet, download the file from the \href{https://github.com/Otcavo/Ipelets-Full-Library}{GitHub} repository, and place it in the \texttt{ipelets} sub-folder of your Ipe folder. This work builds on previous ipelets to add functionality in various contexts \cite{postechDNN_ipelet, IPE, parepally2024ipelets, faber2025french}. Our contribution is in line with that of several other researchers who have worked on neighborhoods of point sets, such as Voronoi diagrams and Delaunay triangulations  \cite{postechDNN_ipelet, IPE}, and $\beta$-skeletons \cite{beta_skeleton}.  Ipelets can be invoked from the ``Ipelets'' menu, see Figure \ref{fig:instructions} for usage instructions.

\section{Neighborhood graphs}
In this section, we describe the neighborhood graphs our Ipelets compute on point sets. We loosely separate them by type and briefly describe our contributions.

\subsection{Nearest-neighbor variants}
\begin{figure}[htbp]
    \centerline{\includegraphics[scale=0.7]{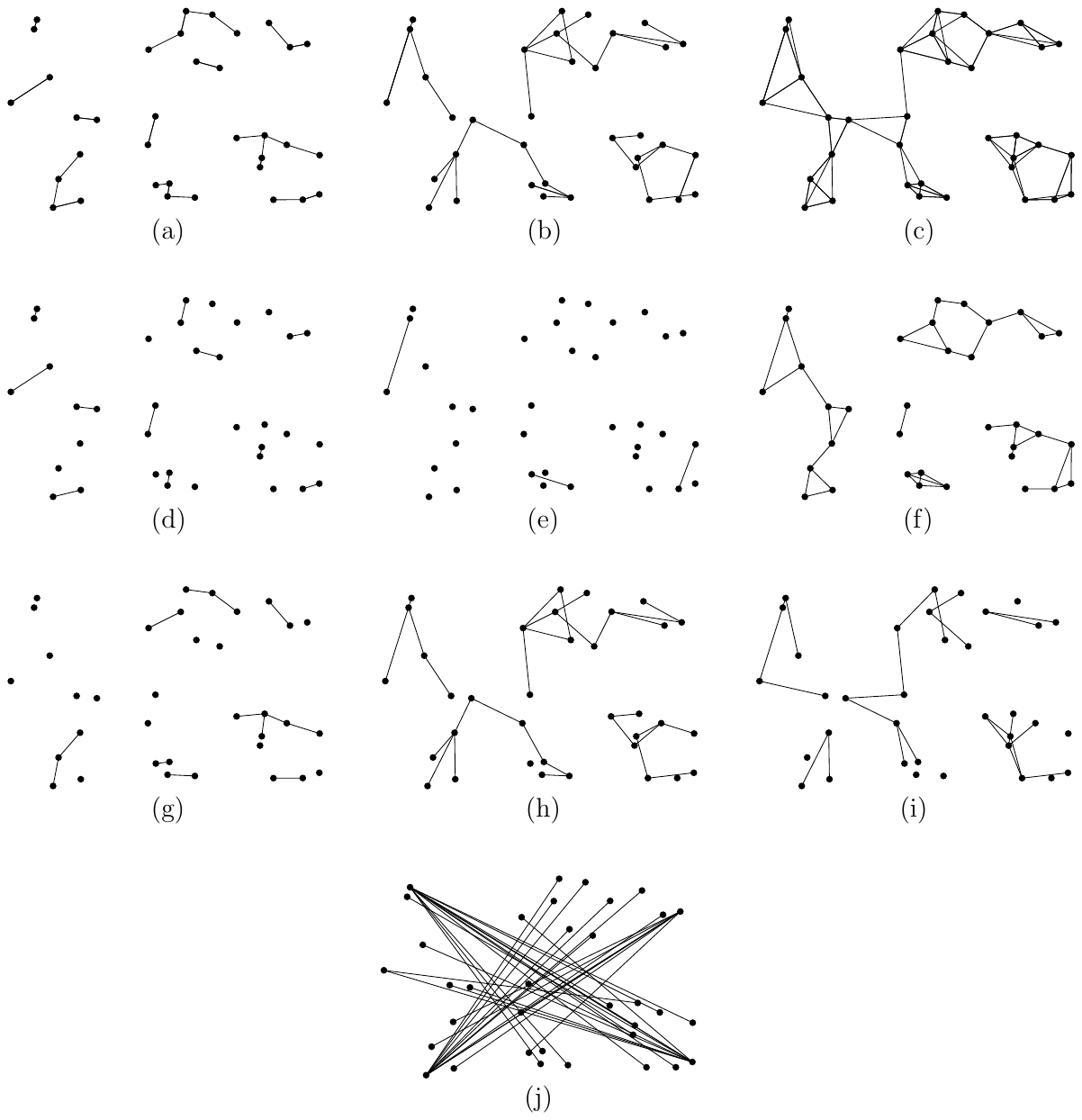}}
    \caption{(a) Nearest, (b) $k^{th}$-nearest, (c) $k$-nearest, (d) mutual nearest, (e) $k^{th}$-mutual, (f) $k$-mutual, (g) asymmetric nearest, (h) asymmetric $k^{th}$-nearest, (i) asymmetric $k$-nearest, and (j) furthest neighbor graphs. In each $k=3$.}
    \label{fig:neighbor}
\end{figure}

\emph{Nearest neighborhood graphs} are fundamental tools in data analysis on a point set $P$. A point $p \in P$ is connected to $q\in P$ when one is the nearest neighbor of the other. The \emph{$k$-nearest neighbor graph} connects $p$ to $q$ when $q$ is among the $k$ nearest neighbors of $p$; the \emph{$k^{th}$-nearest neighbor graph} connects $p$ to $q$ when $q$ is the $k^{th}$ nearest neighbor of $p$; \emph{mutual nearest neighbor graphs} connect $p$ to $q$ when each is the other's nearest neighbor; the \emph{$k$-mutual neighbor graph} connects $p$ to $q$ when each is among the other's $k$ nearest neighbors; the \emph{$k^{th}$-mutual neighbor graph} connects $p$ to $q$ when each is exactly the other's $k^{th}$ nearest neighbor; \emph{asymmetric nearest neighbor graphs} connect $p$ to $q$ when exactly one is the other's nearest neighbor; \emph{asymmetric $k$-nearest neighbor graphs} connect $p$ to $q$ when exactly one has the other among their $k$ nearest neighbors; \emph{asymmetric $k^{th}$-nearest neighbor graphs} connect $p$ to $q$ when exactly one has the other as their $k^{th}$ nearest neighbor; and \emph{furthest neighbor graphs} connect $p$ to $q$ when $q$ is $p$'s furthest neighbor. See illustration in Figure \ref{fig:neighbor}.

\subsection{Proximity graphs}

\begin{figure}[htbp]
    \centerline{\includegraphics[scale=0.7]{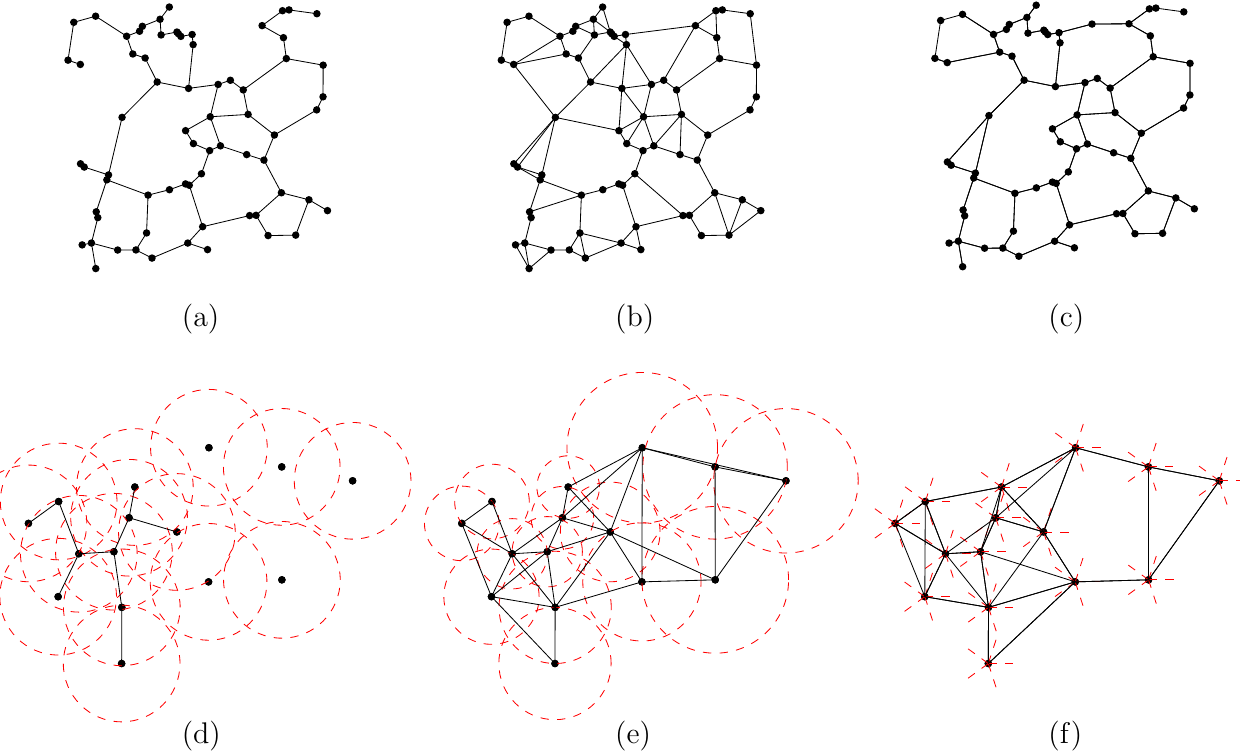}}
    \caption{(a) Relative neighbor, (b) Gabriel graph, (c) Urquhart graph, (d) $\eps$-neighbor ($\eps = 28$), (e) sphere of influence, and (f) Yao graph ($k = 5$).}
    \label{fig:prox}
\end{figure}

A proximity graph is a graph in which points are connected when a region-based condition is satisfied, often in the context of network analysis. The \emph{Gabriel graph} connects two points $p$ and $q$ when the disk defined by the diameter of $p$ and $q$ contains no other points \cite{gabriel_graph}. A \emph{relative neighborhood graph} connects two points $p$ and $q$ exactly when no point is closer to both $p$ and $q$ than they are to each other, or equivalently, no point is in the lune of $p$ and $q$ (intersection of the ball at $p$ with $q$ on its boundary and the ball at $q$ with $p$ on its boundary) \cite{relative_neighborhood_graph}. A \emph{sphere of influence graph} assigns to each point a disk with radius given as the distance to its nearest neighbor and connects points when their disks overlap \cite{sphere_of_influence}. An \emph{$\eps$-graph} connects pairs of points if they lie within $\eps$ of each other. A \emph{Urquhart graph} is the Delaunay triangulation with the longest edge of each triangle removed\cite{urquhart_graph}. A \emph{Yao graph} divides the plane into sectors by angle around points, and points are connected to their closest neighbor per sector\cite{yao_graph}. See visuals in Figure \ref{fig:prox}.

\section{Clustering}
In this section, we describe the clustering algorithms we implemented for point sets. As before, we loosely separate them by type and briefly describe our contributions. 

\subsection{Centroid-based}
\begin{figure}[htbp]
    \centerline{\includegraphics[scale=0.7]{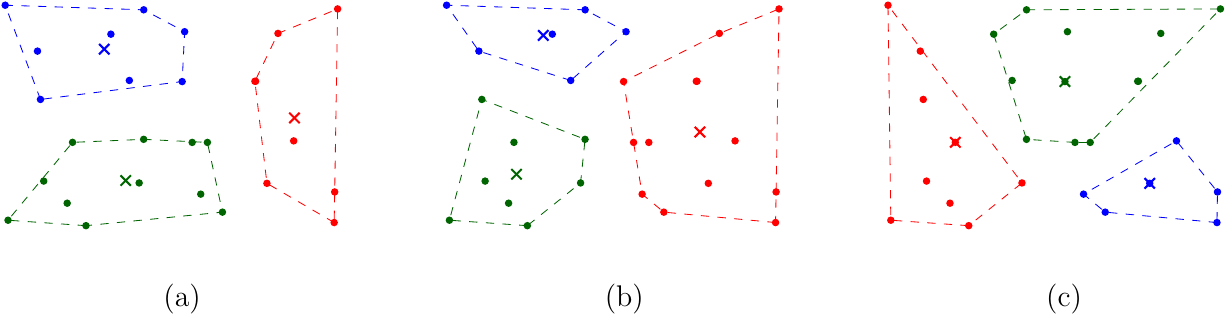}}
    \caption{(a) $k$-Means, (b) $k$-Means++, and (c) $k$-Medoids}
    \label{fig:centroids}
\end{figure}

Centroid-based clustering partitions a point set in space into clusters using a centroid-based method. The \emph{$k$-Means} clustering partitions points into $k$ clusters, assigning points to its nearest centroid and recomputing the centroids iteratively until it is stable. The \emph{$k$-Means++} variant is similar, except it chooses initial centroids that are likely to be far apart to improve convergence. Unlike \emph{$k$-means}, \emph{$k$-Medoids} clustering uses preexisting points in the data as the centroids of the cluster \cite{park2009simple}. Each iteration, a point replaces the medoid if it decreases the total intra-cluster distance. Illustrations of the methods can be seen in Figure \ref{fig:centroids}.

\subsection{Hierarchical}
\begin{figure}[ht]
    \centerline{\includegraphics[scale=0.7]{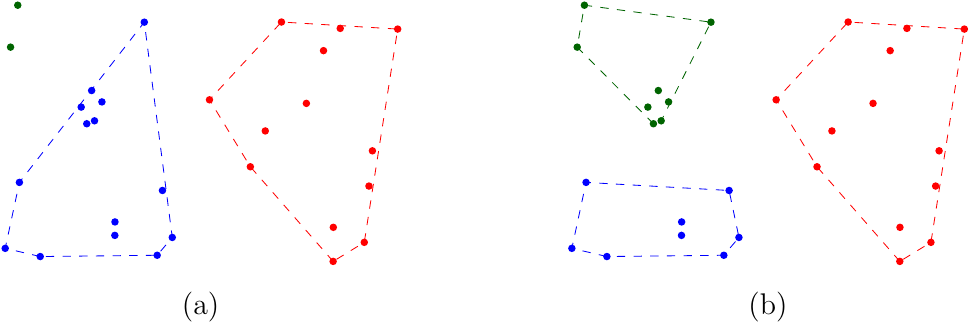}}
    \caption{(a) Single Linkage Clustering, (b) Complete Linkage Clustering}
    \label{fig:hierarchical}
\end{figure}

Hierarchical clustering methods work by treating each point as a cluster and iteratively merging clusters until only a specified number remain. \emph{Single-linkage clustering} defines the cluster distance between clusters as the minimum pairwise distance between their points, while \emph{complete-linkage clustering} uses the maximum \cite{single_complete_linkage}. See Figure \ref{fig:hierarchical}.

\subsection{Density-based}
\begin{figure}[htbp]
    \centerline{\includegraphics[scale=0.7]{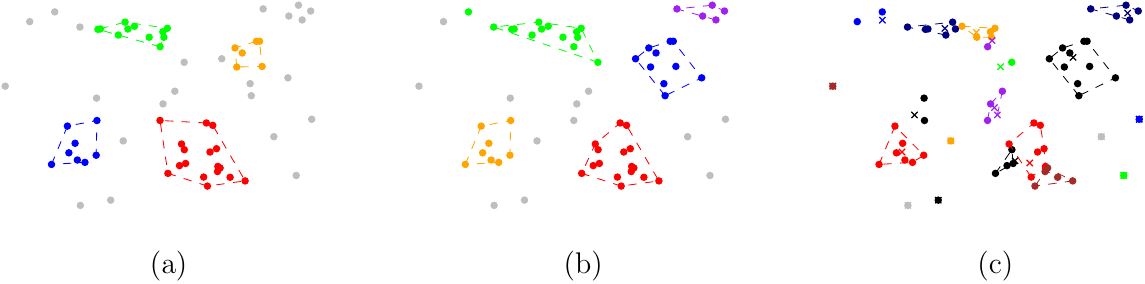}}
    \caption{(a) HDBSCAN, (b) DBSCAN, and (c) Mean shift}
    \label{fig:density}
\end{figure}

Density-based clustering partitions a point set into clusters based on how densely points are packed. \emph{DBSCAN} groups points that are within a specified distance, $\eps$, of each other and have a minimum number of neighboring points (MinPts). Points meeting these criteria form clusters (distinguished by colors), while points that don't are labeled as noise \cite{dbscan}. \emph{HDBSCAN}, on the other hand, applies a hierarchical approach to density-based clustering \cite{hdbscan}. It builds a mutual reachability graph and a minimum spanning tree to identify clusters at varying densities. Instead of relying on a fixed distance threshold like \emph{DBSCAN}, \emph{HDBSCAN} evaluates the stability of clusters across different scales. It then uses the Excess of Mass method to select stable clusters. \emph{Mean shift} computes where each point would converge if moved towards the mean of its neighbors within the given bandwidth radius. Points whose trajectories converge on the same location form a cluster\cite{mean_shift}. See Figure \ref{fig:density}.

\section{Conclusion}
In this paper, we presented a collection of Ipelets for visualizing neighborhood graphs and clustering algorithms on point sets in $\RE^2$. These tools are designed to assist in the understanding of these structures and their properties, and we hope they will be useful to researchers, educators, and artists alike. All Ipelets are freely available on \href{https://github.com/Otcavo/Ipelets-Full-Library}{GitHub}.

\begin{figure}[htbp]
    \centerline{\includegraphics[scale=0.7]{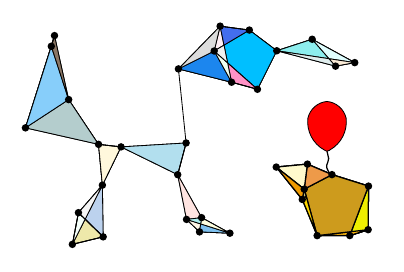}}
    \caption{Artwork made using the $k$ nearest neighbors Ipelet.}
\end{figure}

\bibliography{shortcuts,hilbert}

\end{document}